\documentclass[12pt]{iopart}

\begin{document}
\jl{31}
\title[...solution to the N-body problem]{On the regular-geometric-figure
solution to the N-body problem} 
\author{Antonio S de Castro\footnote[1]{castro@feg.unesp.br} and Cristiane
A Vilela} 
\address{UNESP Campus de Guaratinguet\'a DFQ \\
Caixa Postal 205 \\
12500-000 Guaratinguet\'a SP Brasil}

\begin{abstract}
The regular-geometric-figure solution to the $N$-body problem is presented
in a very simple way. The Newtonian formalism is used without resorting to a
more involved rotating coordinate system. Those configurations occur for
other kinds of interactions beyond the gravitacional ones for some special
values of the parameters of the forces. For the harmonic oscillator, in
particular, it is shown that the $N$-body problem is reduced to $N$ one-body
problems.
\end{abstract}
\submitted
\maketitle

Despite the efforts of mathematicians and physicists over more than two
centuries of research the general problem of $N$ mutually interacting bodies
moving according to Newton's laws, for $N>2$, has never been solved exactly.
The two-body problem subject to forces which depend on the relative vector
positions can be reduced to two one-body problems, one of which describes
the motion of the center of mass and the other one the relative motion. $N=3$
is the smallest value of $N$ that turns the $N$-body problem unsolvable in
the general case. However, under special assumptions about the type of
motion and interaction, analytical solutions of the $N$-body problem can be
found.

In the case of the three-body problem with gravitational interactions some
special solutions are usually presented in the textbooks on classical
mechanics. In the so-called restricted three-body problem two heavy bodies
move about the common center of mass whereas a third light body moves in the
same plane \cite{mar}-\cite{hes}. In the so-called Lagrange's case the three
bodies are at the vertices of an equilateral triangle in every part of the
motion, which rotates about an axis perpendicular to the plane of the bodies
while it changes in size \cite{mar1}-\cite{som}. There is still another
special solution for three bodies interacting by gravitational forces known
as Euler's case. In that last case the bodies move along the same straight
line during all the motion \cite{sym2}-\cite{som2}. Another special solution
is that of $N$ bodies of similar masses subject to similar forces moving on
the vertices of a regular $N$-sided polygon \cite{som3}. All of these
special solutions are of great pedagogical importance since they give
solutions of a problem unsolvable in the general case. Nevertheless the
resolution of Lagrange's case, as presented by textbook authors, resorts to
a rotating coordinate system requiring an extensive calculation, and
consequently to a weakening of the pedagogical attractiveness.

In a previous work \cite{asc2} Lagrange's case has been presented in an
alternative and more general way, permitting that it can be easily
approached immediately after the presentation of the two-body problem. In
that paper the equilateral-triangle solutions were obtained for interactions
which go beyond the gravitational ones. Ess\'{e}n, in a recent paper to this
journal \cite{ess} (homonym to that one in Ref. \cite{asc2}), approached
this very same problem using the Lagrangian formalism restricting himself to
gravitational interactions and presenting an extension to the $N$-body
problem. Encouraged by the results obtained in \cite{ess} and following the
same steps of Ref. \cite{asc2}, in this paper we reduce the $N$-body problem
to $N$ one-body problems in the case of more general central two-body
interactions, such that the total force on each body is directed towards, or
from the center of mass of the $N$-body system. As a by-product, for
gravitational interactions we obtain an extension of the Lagrange case. We
obtain that the $N$ bodies are on the vertices of a non-rigid (in general)
regular geometric figure throughout the motion. For other kind of
interactions, this is also possible if certain conditions are satisfied by
the parameters characterizing the intensities of the forces.  

Let us consider $N$ bodies (treated as particles) with masses $m_{i}$ $%
(i=1,\ldots N)$ located by the vectors $\mathbf{r}_{i}$. It is supposed that
the forces are pairwise, directly proportional to an arbitrary exponent of
the distance between the bodies and directed along the line connecting them.
Thus the resulting force acting on the \textit{i}-th body due to the other $%
N-1$ bodies can be written as

\begin{equation}
\mathbf{F}_{i}=-\sum_{\stackrel{j=1}{(j\not{=}i)}}^{N}K_{ij}\frac{\mathbf{r}%
_{i}-\mathbf{r}_{j}}{r_{ij}^{n}}  \label{eq1}
\end{equation}

\noindent where $K_{ij}$ $(K_{ii}=0$ and $K_{ij}=K_{ji}$, according to
Newton's third law$)$ is the proportionality constant, $r_{ij}$ is the
distance between the bodies $i$ and $j$, and the exponent $n$ is a real
number. Included within the possibilities allowed by (\ref{eq1}) are some
familiar forms of interactions such as the gravitational and the harmonic
oscillation ones. For the gravitational case we have $n=3$ and $%
K_{ij}=Gm_{i}m_{j}$, where $G$ is the universal gravitation constant,
whereas for a harmonic-oscillator-type interaction $n=0$.

The $N$-body problem is not subject to external forces and consequently the
center of mass of the system has no acceleration and, for the sake of
simplicity, it is considered as being in rest. The position of the \textit{i}%
-th body from the center of mass of the system $\mathbf{r}_{i}^{\prime }$ is
related to $\mathbf{r}_{i}$ by

\begin{equation}
\mathbf{r}_{i}=\mathbf{r}_{i}^{\prime }+\mathbf{R}  \label{eq2}
\end{equation}

\noindent where $\mathbf{R}$ is the vector position of the center of mass.
Therefore in the center-of-mass system frame the force acting on the \textit{%
i}-th body takes the form

\begin{equation}
\mathbf{F}_{i}=-\sum_{\stackrel{j=1}{(j\not{=}i)}}^{N}\frac{K_{ij}}{%
r_{ij}^{n}}\mathbf{r}_{i}^{\prime }+\sum_{\stackrel{j=1}{(j\not{=}i)}}^{N}%
\frac{K_{ij}}{r_{ij}^{n}}\mathbf{r}_{j}^{\prime }  \label{eq3}
\end{equation}

\noindent This force can also be written in the form

\begin{equation}
\mathbf{F}_{i}=-\sum_{\stackrel{j=1}{(j\not{=}i)}}^{N}\frac{K_{ij}}{%
r_{ij}^{n}}\mathbf{r}_{i}^{\prime }+\sum_{\stackrel{j=1}{\stackrel{(j\not%
{=}i)}{(j\not{=}k)}}}^{N}\frac{K_{ij}}{r_{ij}^{n}}\mathbf{r}_{j}^{\prime }+%
\frac{K_{ik}}{r_{ik}^{n}}\mathbf{r}_{k}^{\prime }\qquad (k\ne i)  \label{eq4}
\end{equation}

\noindent Using the fact that

\begin{equation}
\sum_{i=1}^{N}m_{i}\mathbf{r}_{i}^{\prime }=\mathbf{0}  \label{eq5}
\end{equation}

\noindent by the definition of the center of mass frame, we can write

\begin{equation}
\mathbf{r}_{k}^{\prime }=-\frac{1}{m_{k}}\sum_{\stackrel{s=1}{(s\not{=}k)}%
}^{N}m_{s}\mathbf{r}_{s}^{\prime }  \label{eq6}
\end{equation}

\noindent After substituting (\ref{eq6}) in the last term of (\ref{eq4}) and
rearranging all the terms we get

\begin{equation}
\mathbf{F}_{i}=-\left( \sum_{\stackrel{j=1}{(j\not{=}i)}}^{N}\frac{K_{ij}}{%
r_{ij}^{n}}+\frac{K_{ik}}{r_{ik}^{n}}\frac{m_{i}}{m_{k}}\right) \mathbf{r}%
_{i}^{\prime }+\sum_{\stackrel{j=1}{\stackrel{(j\not{=}i)}{(j\not{=}k)}}%
}^{N}\left( \frac{K_{ij}}{r_{ij}^{n}}-\frac{K_{ik}}{r_{ik}^{n}}\frac{m_{j}}{%
m_{k}}\right) \mathbf{r}_{j}^{\prime }\quad (k\ne i)  \label{eq7}
\end{equation}

\noindent If the following condition is satisfied

\begin{equation}
\frac{\frac{K_{ij}}{r_{ij}^{n}}}{m_{i}m_{j}}=
\frac{\frac{K_{ik}}{r_{ik}^{n}}}{m_{i}m_{k}}\qquad (j\ne i,k\ne i)\label{eq8}
\end{equation}

\noindent {\it i.e.},

\begin{equation}
\frac{K_{ij}}{r_{ij}^{n}}=\lambda m_{i}m_{j} \label{eq8a}
\end{equation}

\noindent for some proportionality factor $\lambda$ characteristic of the
interactions as well as the geometrical configuration of the system, the
resulting force acting 
on each particle will be along the line connecting it to the common center
of mass of the $N$-body system. When the parameters $K_{ij}$ and masses meet
the requirement of (\ref{eq8}) we have

\begin{equation}
\mathbf{F}_{i}=-\frac{K_{ij}^{e\!f\!f}}{r_{ij}^{n}}\mathbf{r}_{i}^{\prime
}\qquad (j\ne i)  \label{eq9}
\end{equation}

\noindent where

\begin{equation}
K_{ij}^{e\!f\!f}=\lambda m m_{i}  \label{eq10}
\end{equation}

\noindent is the effective proportionality constant and $m=%
\sum_{i=1}^{N}m_{i}$ is the total mass of the system. Therefore the $N$-body
problem reduces to $N$ coupled problems of bodies exposed to central forces,
with the common center of force located in the center of mass of the system.
The coupling occurs by the presence of $r_{ij}$ in (\ref{eq9}).

As a particular case one has the gravitational one, for the three-body
problem with the three bodies moving on the vertices of an equilateral
triangle and in addition for four bodies on the vertices of a regular
tetrahedron ($r_{ij}=r_{ik}$, for all $i$, $j$ and $k$ with $i\neq j$ and $%
i\neq k$). One should also note that there is no reason for maintaining $%
r_{ij}=const$, \textit{i.e.}, a rigid regular geometric figure, it is just
enough that any side or edge be congruent to any other during all the time
in order to maintain the regular character of the geometric figure. Thus, it 
means that $\lambda $ may vary with time. In other
words, the regular geometric figures may change in size with their sides or
edges expanding or contracting at the same rhythm, as in the original
Lagrange's case (three gravitating bodies, $r_{ij}^3=G/\lambda$), and in the 
case of the
tetrahedron as well. As a matter of fact the equilateral-triangle figure may
also remain rigid while the bodies rotate about an axis perpendicular to the
plane with a common angular velocity, and in this situation the bodies move
on circles centered at the center of mass of the system, which coincides
with the common center of force. For a regular-tetrahedral figure things are
indeed different because one should have four bodies on the vertices of a
rigid tetrahedron moving on the great circles of concentric spheres, and
this is impossible even in the case of equal masses when one should have
just one sphere.

The equilateral-triangle and the regular-tetrahedral solutions also occur in
a more general situation where

\begin{equation}
\frac{K_{ij}}{K_{ik}}=\frac{m_{j}}{m_{k}}  \label{eq11}
\end{equation}

\noindent an $n$-independent condition, equivalent to $K_{ij}=%
\alpha_{i}m_{j} $, where $\alpha _{i}$ is a constant.

It is worthwhile to note that the constraint (\ref{eq11}) implies that the $%
N $-body problem reduces to $N$ one-body problems only in the case of the
harmonic-oscillator-type interactions ($n=0$), here the geometric links are
missing in (\ref{eq8}) and (\ref{eq9}) thus each of the $N$ bodies moves
independently of the others as if they were bounded to the center of mass of
the $N$-body system by springs of elastic constants given by $%
K_{ij}^{e\!f\!f}=\alpha _{i}m$, so they do not need to be on the vertices of
a regular geometric figure. The condition (\ref{eq11}) has the same form as
that one found in Ref. \cite{asc} showing the necessary condition for the
Jacobi coordinates ${\rho }=\mathbf{r}_{2}-\mathbf{r}_{1}$ and ${\lambda }=%
\mathbf{r}_{3}-\frac{m_{1}\mathbf{r}_{1}+m_{2}\mathbf{r}_{2}}{m_{1}+m_{2}}$
be conducive to the separation of the variables in the three-body problem
with harmonic interactions.

It is worth pointing out that other solutions to the $N$-body problem with
central forces can be obtained if the condition (\ref{eq8}) is not imposed
for all $i$, $j$ and $k$ but only for $j$ and $k$ which label the nearest
particles of \textit{i}-th one. This last condition is less restrictive than
that one presented in Ref. \cite{som} which requires $N$ bodies of similar
masses subject to similar forces moving on the vertices of a regular $N$%
-sided polygon. Taking into account that weaker condition one is conduced to
all sorts of regular geometric figures. Regular polygons can rotate about an
axis perpendicular to the plane with a common angular velocity while they
change their sides at the same pace. For regular polyhedrons, though,
expansions or contractions are the only permitted movements.

\vspace{0.5in}

\ack{The authors would like to thank an anonymous referee for useful
suggestions 
 and the FAPESP for partial support.}


\begin{thebibliography}{99}
\bibitem{sym2}  This case is left as an exercise in Ref. 2, Sec. 12.8, p 494.
\bibitem{som2}  The existence of this solution  is mentioned in Ref. 7,
Chap. V, ~\S\ 32, p 180
\end{thebibliography}
\end{document}